# Probabilistic Diffusion in Random Network Graphs

Dr. Natarajan Meghanathan
Professor of Computer Science
Jackson State University, MS, USA
E-mail: natarajan.meghanathan@jsums.edu

**Abstract**

In this paper, we consider a random network such that there could be a link between any two nodes in the network with a certain probability ($p_{link}$). Diffusion is the phenomenon of spreading information throughout the network, starting from one or more initial set of nodes (called the early adopters). Information spreads along the links with a certain probability ($p_{diff}$). Diffusion happens in rounds with the first round involving the early adopters. The nodes that receive the information for the first time are said to be covered and become candidates for diffusion in the subsequent round. Diffusion continues until all the nodes in the network have received the information (successful diffusion) or there are no more candidate nodes to spread the information but one or more nodes are yet to receive the information (diffusion failure). On the basis of exhaustive simulations conducted in this paper, we observe that for a given $p_{link}$ and $p_{diff}$ values, the fraction of successful diffusion attempts does not appreciably change with increase in the number of early adopters; whereas, the average number of rounds per successful diffusion attempt decreases with increase in the number of early adopters. The invariant nature of the fraction of successful diffusion attempts with increase in the number of early adopters for a random network (for fixed $p_{link}$ and $p_{diff}$ values) is an interesting and noteworthy observation (for further research) and it has not been hitherto reported in the literature.

## 1 Introduction

We envision a random network of nodes such that information could propagate along any link with a certain probability. We are interested in the problem of analyzing how fast information (for example, the availability of funds, need for help, etc) originating from one or more nodes (referred to as the early adopters) in the network can diffuse (spread) to the other nodes of the network. Note that diffusion may not happen across all the links because there is only a certain chance with which a node may share the information to its neighbor node. Though probabilistic diffusion has been widely studied in the area of complex network analysis, most of the studies (e.g., [3, 6, 8]) are restricted to real-world network models and not conducted on theoretical models such as those that correspond to the random networks. Our conjecture is that phenomenon observed in random network models could be construed as those that simply happen by chance (due to the degree distribution of the vertices and not due to the nodes involved). If a similar phenomenon is observed in a real-world network whose degree distribution is similar to that of a random network, then we could conclude that the phenomenon observed in the real-world network also simply happens due to the distribution of the vertices and not due to the specific nature of the nodes involved. The above characteristic of random networks forms the motivation for our research in this paper. We are interested in analyzing the impact of increase in the number of early adopters on the success of diffusion in a random network (i.e., whether all the nodes in the network receive the information) and the delay associated with a successful diffusion attempt.

The rest of the paper is organized as follows: Section 2 presents the system model and explains the diffusion phenomenon in random network graphs with an example. Section 3 presents the algorithm to construct random networks and presents the results of the simulation analyzing the impact of the number of early adopters on diffusion in a random network. Section 4 discusses related work. Section 5 draws conclusions. Throughout the paper, we use the terms 'node' and 'vertex' as well as 'link' and 'edge' interchangeably. They mean the same.

## 2 Diffusion

We assume a network graph $G = (V, E)$ where $V$ is the set of vertices and $E$ is the set of edges. Let $p_{link}$ represent the probability of link between any two vertices $u$ and $v$ (suits the random graph model, for more details, see Section 3). The edges are undirected. The degree of a vertex is the number of vertices adjacent to the vertex in the graph. For random graphs, the degrees of the vertices are comparable to each other (see Figure 2). The graph is considered to be connected if we can reach from any vertex to any other vertex through one or more hops. We test for connectivity of a graph using the well-known Breadth First Search algorithm [5]. We consider only connected graphs for diffusion.

Diffusion is the process of spread of information originating from one or more vertices (called the early adopters) to the rest of the vertices in the graph [11]. Each vertex attempts to spread the information received from one of its adjacent vertices to the other adjacent vertices. Let $p_{diff}$ be the probability for a vertex $u$ to disseminate the information to spread to its neighbor vertex $v$, and is the same for every edge. A vertex attempts to spread the information only when it receives the information for the first time (to avoid looping of the information). When a vertex $u$ attempts to spread the information to a vertex $v$, we generate a random number $r_{u->v}$ in the range [0…1] and if $r_{u->v} \leq p_{diff}$, then the information is passed on from $u$ to $v$, otherwise not. We maintain a set of nodes called the covered nodes that have received the information from at least one of their neighbors across the rounds of the diffusion process. We also maintain a candidate set of nodes, updated with every round of diffusion (more details below).

To start with, the set of early adopters are assumed to be the set of covered nodes as well as the set of candidate nodes. Diffusion proceeds in rounds. In each round, each vertex in the candidate set of nodes attempt to spread the information they have received (in the previous round or the initialization stage, in the case of the early adopters) to each of their neighbors. The neighbor nodes that receive the information for the first time are added to the set of covered nodes and are also added to the set of candidate nodes for diffusion in the subsequent round. The set of candidate nodes is refreshed during each round. A node could get into the candidate set of nodes for diffusion in the next round only if the node has been covered for the first time and has not attempted to spread the information until then. We consider the diffusion process to be successful for the entire network if the candidate set of nodes for the next round of diffusion becomes empty and all the nodes in the network are covered by then (i.e., all the nodes have received the information at least once). A diffusion process is considered to be unsuccessful if one or more nodes in the network are yet to receive the information and the set of candidate nodes for the next round of diffusion gets empty.

We now explain the diffusion process using an example shown in Figures 1 and 2. The input graph used in both the figures is the same: the number inside the circle is the node ID; though the input graph is an undirected graph of edges – we generate two different random numbers for each edge, one for each direction. The probability for diffusion ($p_{diff}$) along any edge is assumed to be 0.50. There could be diffusion from node $u$ to node $v$ only if the random number assigned for the edge $(u, v)$ in the direction $u->v$ is less than or equal to $p_{diff}$. Accordingly, we generate an initial graph for the input graph as shown in Figures 1 and 2. The two figures differ in the choice of the early adopter node used to initiate diffusion and the resulting sequence of rounds. Diffusion proceeds in rounds – for each round, the candidate set of nodes are colored in blue and the nodes covered across all the rounds are colored in yellow. Figure 1 illustrates a successful diffusion starting from node 0 (the early adopter node) and it takes a total of 4 rounds for diffusion to successfully complete, with the following nodes forming the candidate set for each round: round 1 (node 0), round 2 (nodes 1 and 3), round 3 (node 4) and round 4 (node 5). Figure 2 illustrates an unsuccessful diffusion starting from node 4 (the early adopter node) and proceeding up to 3 rounds (the candidate set of nodes are - round 1: node 4; round 2: node 1 and node 5; round 3: node 2) after which there is no scope for further diffusion (as the candidate set of nodes for round 4 is empty), but nodes 0 and 3 are yet to be covered.

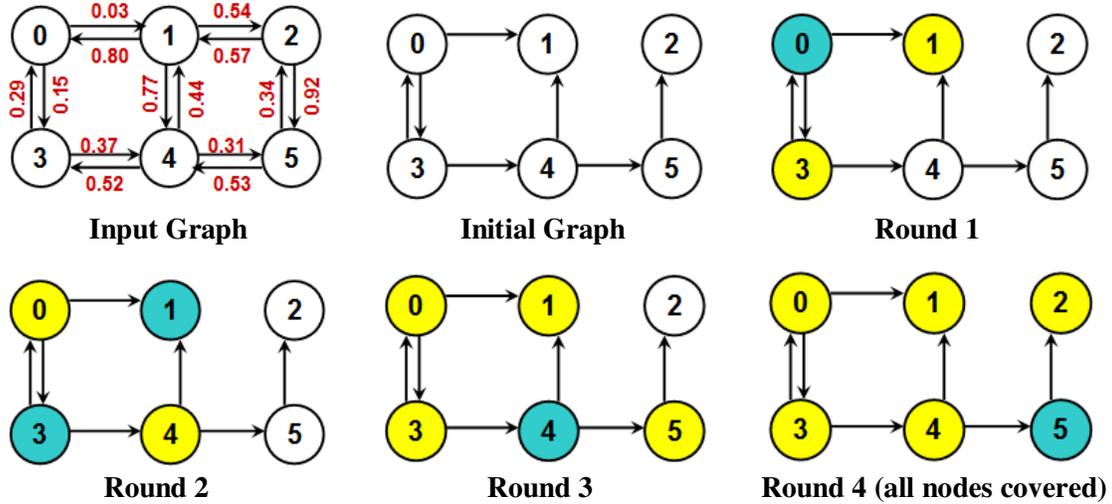

**Figure 1:** Example for a Successful Diffusion (Probability of Diffusion on an Edge, $p_{diff}$ = 0.50)

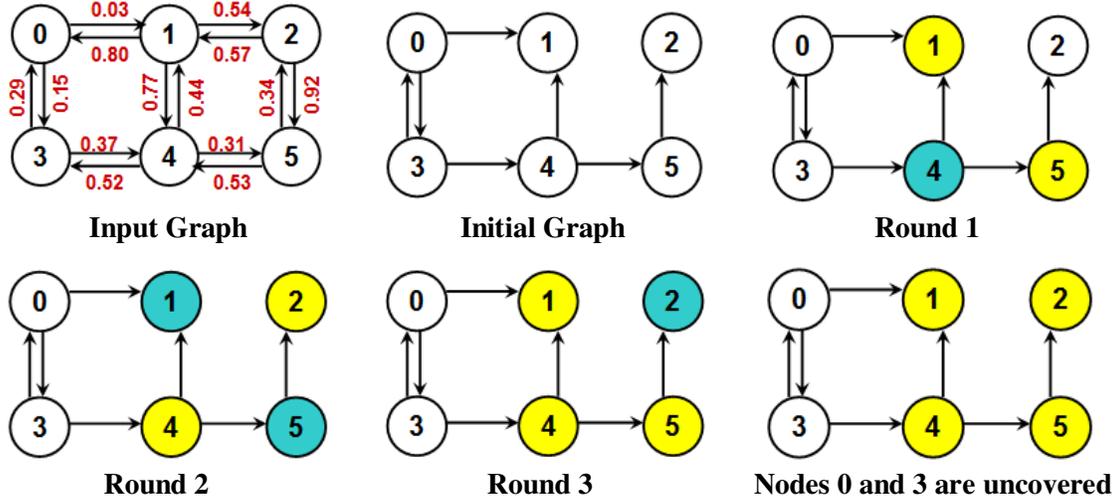

**Figure 2:** Example for an Unsuccessful Diffusion (Probability of Diffusion on an Edge, $p_{diff}$ = 0.50)

## 3 Random Graphs and Simulations

We use the well-known Erdos-Renyi model [2] to generate the random graphs for the simulations. The model takes as inputs - the number of nodes (*N*) in the network and the probability of a link ($p_{link}$) between any two nodes in the network. For any two pair of nodes *u* and *v* (where *u* < *v*), we generate a random number $r_{u-v}$ and if $r_{u-v} \leq p_{link}$, we set up an undirected link between *u* and *v* in the network. The larger the value of $p_{link}$ and/or the larger the total number of nodes in the network, the more dense is the network as well as the more closer are the degrees of the vertices to the average node degree (observed based on the reduction in the standard deviation of the node degrees with increase in the $p_{link}$ values and/or with increase in the total number of nodes; see Figure 3). We conduct simulations with 100 nodes and 200 nodes; the $p_{link}$ values used are 0.05, 0.10, 0.15, 0.20 and 0.30; the $p_{diff}$ values used are 0.05 to 1.0, in increments of 0.05; the values for the number of early adopters are 1, 10 and 20. We run 200 instants of the simulations for each combination of values for the above parameters (total # nodes, $p_{link}$, $p_{diff}$ and # early adopters) and average the results to measure the following two metrics (95% confidence interval): (i) Probability of successful diffusion and (ii) Average number of rounds per successful diffusion attempt. For each combination of values for the above parameters, the probability of successful diffusion is the number of simulation runs leading to a successful diffusion divided by the total number of simulation runs (which is 200 runs); for each such successful diffusion attempt, we count the number of rounds it takes for the information originating from one or more early adopters to reach all the nodes in the network and average the values for the number of rounds across all the successful diffusion attempts.

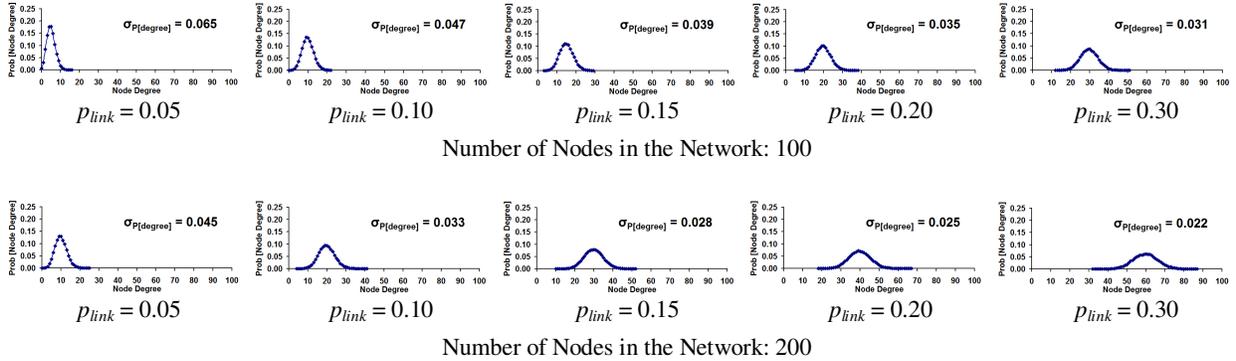

**Figure 3:** Degree Distribution for a Random Network Graph and the Variation in Node Degrees

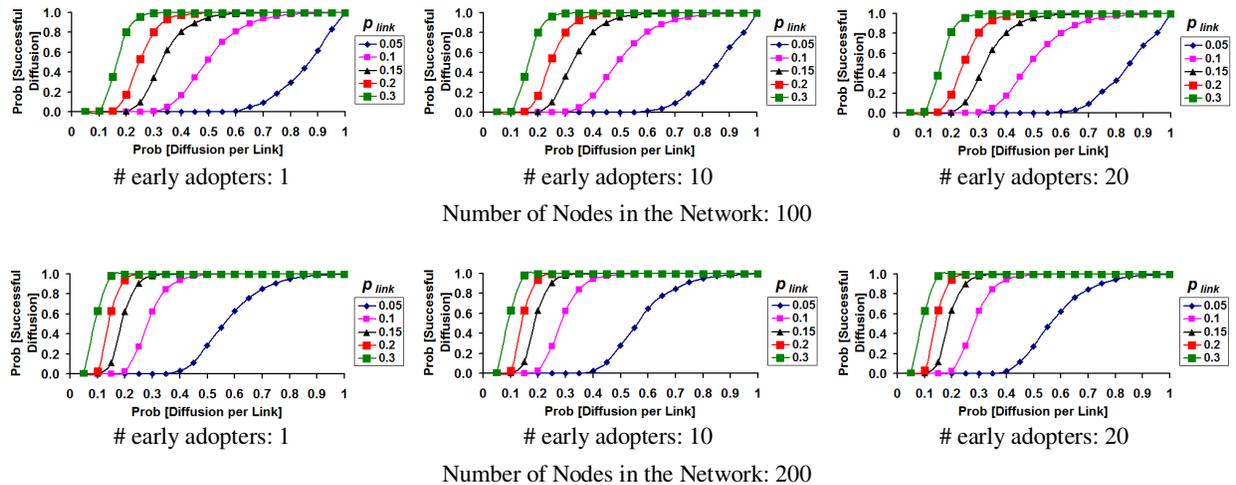

**Figure 4:** Probability of Successful Diffusion for a Random Network Graph vs. # Early Adopters

The most interesting and significant observation from the simulation results is with regards to the probability of successful diffusion for the three different values for the number of early adopters for a given number of nodes in the random network graph. Though for a given $p_{link}$ value and number of nodes, the probability of successful diffusion increases with increase in the $p_{diff}$ value, the nature of increase remains the same, irrespective of the values for the number of early adopters. For a given $p_{link}$, $p_{diff}$ and number of early adopters, we observe the probability for a successful diffusion to increase with increase in the total number of nodes in the network. Except for the $p_{link}$ value of 0.05 for 100 nodes random network, the nature of increase in the probability of successful diffusion for a given $p_{link}$ and number of nodes in the random network increases in a "concave down increasing pattern" with increase in the $p_{diff}$ value. For a given $p_{diff}$, we also observe the probability of successful diffusion to significantly increase (more than an exponential increase) as we increase the $p_{link}$ values in increments of 0.05, especially for $p_{link}$ values of 0.15 and above. After a while ($p_{link}$ values of 0.30 or above), we observe the increase in the probability for successful diffusion to saturate and hence we do not present the simulation results for $p_{link}$ values above 0.30.

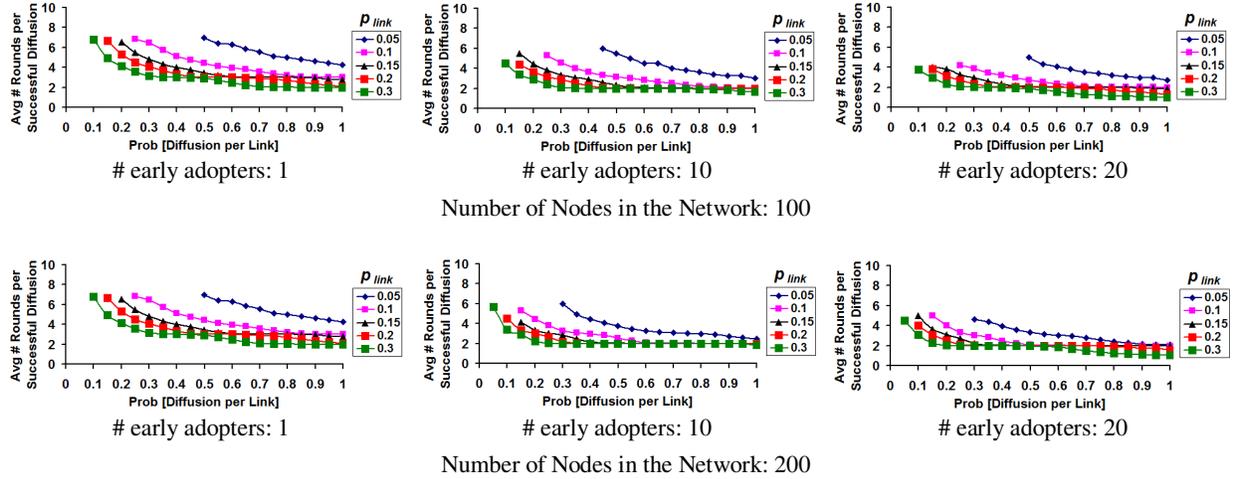

**Figure 5:** # Successful Rounds per Diffusion for a Random Network Graph vs. # Early Adopters

With regards to the average number of rounds per successful diffusion attempt, for a given number of nodes in the random network, # of early adopters and $p_{link}$ value, we observe the metric to exhibit a "concave up, decreasing" pattern of decrease with increase in the $p_{diff}$ values. Though as expected, for a given $p_{diff}$, $p_{link}$ and the total number of nodes in the network, the average number of rounds per successful diffusion attempt decreases with increase in the number of early adopters, the nature and magnitude of decrease is not as high as the decrease with increase in $p_{diff}$ (for a given $p_{link}$, initial # responders and the total number of nodes). Even for a lower number of early adopters, we observe the average number of rounds per successful diffusion to saturate (to the lowest value incurred for a particular value for the total number of nodes) for $p_{diff}$ values around 0.30, indicating that it may not be necessary to operate with a significantly larger number of early adopters to decrease the average number of rounds per diffusion in a random network. We do not report the average number of successful rounds per diffusion for $p_{link}$ and $p_{diff}$ values that do not incur any successful diffusion in the simulation runs.

## 4  Related Work

The contagion model [9] has been the most commonly used model for diffusion in complex networks. According to this model, given two choices of behaviors (say A or B), the early adopters are considered to choose one of the two behaviors (say A), while the rest of the nodes choose the other behavior (say B). Diffusion spreads in rounds, wherein each round, a node decides to change its behavior if a majority of its neighbors (a typical value for the degree-based diffusion threshold) have a behavior different from itself. The contagion model is more relevant for networks with scale-free form of degree distribution and the early adopters are typically nodes with larger degree [10]. The diffusion model considered in this paper is different from the contagion model and is more applicable for networks in which the degree distribution is normal (see Figure 3). We consider diffusion to happen with a certain probability on any link; this way, nodes that are not well-known to each other may still have some form of association between them and be willing to spread the information with a certain probability. While the contagion model required the initial responders to be a non-negligible fraction of the total number of nodes in the network for a successful network-wide diffusion, we observe the probabilistic diffusion model for random networks to be independent of the number of initial responders for successful diffusion.

The probabilistic diffusion model considered in this paper is also different from that of the SIS (susceptible-infected-susceptible) and SIR (susceptible-infected-removed) models for diffusion. Even though both the SIS and SIR models [1] are probabilistic models for diffusion, the *infected* nodes could again change their state (to either *susceptible*, as in the SIS model or *removed*, as in the SIR model); our probabilistic model of diffusion uses the infected nodes to become the candidate nodes for diffusion in the subsequent round and these nodes continue to stay infected throughout the network lifetime. The SIS and

SIR models have been also observed to be dependent on the number of early adopters on the network they are applied on [4].

In [6], the authors propose a probabilistic approach of social influence diffusion model with incentives (as uniform diffusion has been observed to be no longer valid in social networks and high degree nodes need not be the most influential in all contexts [12]); the authors propose an influence diffusion probability for each node, instead of uniform probability, and categorize nodes into two classes: active and inactive; the active nodes have chances of influencing the inactive nodes, but not vice-versa; diffusion still happens based on a system-wide threshold. Our probabilistic diffusion model is link-based (could be even run with different diffusion probability for each link) and does not use any node-based system-wide threshold to regulate the diffusion.

To the best of our knowledge, ours is the first such probabilistic model of diffusion proposed for random networks for which there exists a link between any two nodes with a certain probability; diffusion happens across each link with a certain probability and without the use of a diffusion threshold (that depends directly or indirectly on the degree of the nodes, as in most of the previous works). The observation that under the above probabilistic diffusion model, "the probability for a successful diffusion in a random network does not depend on the number of initial responders," has been hitherto not reported in the literature.

## 5   Conclusions

The high-level contribution of this paper is the application of probabilistic diffusion on random network graphs and the observation from the simulation results that for a given random network and probability of diffusion on a link, the probability for successful diffusion does not depend on the number of early adopters. We also observe that for moderate-larger values of probability of diffusion on a link in a random network, it may not be necessary to operate with a larger number of early adopters to decrease the average number of successful rounds per diffusion attempt. The results observed in this paper are different from the results observed for the contagion as well as the SIS and SIR diffusion models – all of which report that there exists a threshold number of early adopters needed for a successful diffusion for complex networks. The results presented in this paper indicate that at least for random networks, a probabilistic diffusion model – like the one described in this paper – could lead to successful diffusion that is independent of the number of early adopters. We opine that the research presented through this paper could pave way for further studies on probabilistic diffusion in random network graphs and other forms of complex network graphs as well as for real-world network graphs.